# Understanding User Perception and Intention to Use Smart Home for Energy Efficiency: A Survey

Alona Zharova*[1] and Hee-Eun Lee[1]

[1] Humboldt-Universität zu Berlin
Unter den Linden, 6, 10099 Berlin, Germany

* - corresponding author (alona.zharova@hu-berlin.de)

## Abstract

Smart Home technology is increasingly seen as a solution for improving household energy efficiency. However, its energy-saving potential depends largely on how consumers use the system. To explore how user perception and intention to use Smart Home can influence energy efficiency, we develop a research model combining the theory of planned behavior (TPB) and the norm activation model (NAM), based on a comprehensive literature review. We collect data by surveying users of Smart Home systems (N = 363) and apply a partial least squares structural equation model (PLS-SEM) extended by a Random Forest algorithm to capture both linear and non-linear causal relationships. Results show that personal norms, shaped by a sense of responsibility and awareness of environmental consequences, are the strongest predictors of energy-efficient smart home use. Social norms and attitudes also significantly contribute to the intention to use these systems efficiently. Moreover, past behavior strengthens the link between personal norms and behavioral intention, highlighting the role of habit in shaping energy-related actions. To maximize the energy-saving potential of Smart Homes, system design should focus on reinforcing personal moral norms, supporting long-term engagement through habit-forming features, delivering personalized feedback on environmental and financial outcomes, and embedding green automation defaults. Implementing policy mechanisms that financially reward household energy savings presents a powerful lever for reducing emissions through improved energy efficiency in residential buildings.

**Keywords**: Smart Home, energy efficiency, partial least squares structural equation modeling, random forest, behavioral intention, theory of planned behavior, norm-activation model

# 1 Introduction

Mitigating climate change requires deep and rapid reductions in energy-related emissions, particularly in the residential building sector, which accounts for roughly 21% of global greenhouse gas emissions (IPCC, 2022). Improving energy efficiency in buildings is a critical lever for reducing emissions, lowering energy demand, and meeting climate targets (Creutzig et al., 2022). Energy consumption in a building is influenced by various factors, such as climate, characteristics of the building, building service systems and operations, social and economic factors, indoor environmental quality, and user behavior (Abrahamse & Schuitema, 2020; Wan et al. 2025; Steemers & Yun, 2009). Beyond structural retrofits and building code improvements, demand-side digitalization through technologies such as Smart Home systems has emerged as a promising tool for decarbonizing household energy use (Martin & Larsen, 2024; Casquero-Modrego et al., 2025).

Smart Home technologies (SHT) are on the cusp of revolutionizing residential energy management by enabling the automated monitoring and control of household energy consumption (Saad-al-sumati et al., 2014; Sanguinetti et al., 2018). These systems have the potential to reduce both overall consumption and peak demand (Alaa et al., 2017; Lashkari et al., 2019). Various features such as programmable thermostats, energy consumption analytics, or automated lighting systems directly contribute to reducing unnecessary energy usage and enhancing overall efficiency (Risteska Stojkoska & Trivodaliev, 2017; Sun & Li, 2021). Using SHT to shift the load of appliances usage to the times with more renewable energy in the grid is a further way to reduce carbon emissions in residential households (Zharova et al, 2024a; Zharova et al, 2024b). While the technical aspects of SHT and their potential to optimize energy usage have been thoroughly studied (Ford et al., 2017; Ji & Chan, 2019), there is a lack of understanding of the user perspective (Marikyan et al., 2019; Sintov & Schultz, 2017).

User behavior vastly influences how much energy is used (Schwartz et al., 2015). Behavioral changes among building occupants can help reduce GHG emissions in the buildings sector by up to 15% (IPCC, 2022). However, the realization of the energy-saving potential of SH depends on how individuals perceive, engage with, and incorporate SH systems into daily life (Fornara et al., 2025; Peacock et al., 2017). This also includes the consideration of the dynamic nature of household structures which affects the total energy consumption (Irwin, 2017). The positive impact of SH on energy efficiency is further dependent on how consumers are using them after adoption (Hargreaves et al., 2018; Thunshirn et al., 2025; Pakravan & MacCarty, 2020).

In this paper, we develop and empirically test a behaviorally grounded research model that integrates the Theory of Planned Behavior (TPB) and the Norm Activation Model (NAM). Using survey data from 363 Smart Home users, we apply a partial least squares structural equation model (PLS-SEM), complemented by a random forest algorithm, to uncover both linear and non-linear drivers of energy-efficient smart home behavior. Our findings reveal that personal norms shaped by a sense of responsibility and awareness of environmental consequences are the strongest predictors of intention to use smart home systems for energy efficiency. Social influences and attitudes also play a significant role, while perceived behavioral control is less consequential. These results highlight the importance of aligning system design and policy interventions with users' values and motivations. Unlocking the full energy-saving potential of smart homes will require a shift from usability-focused strategies to those that engage moral norms, support habit formation, and deliver personalized environmental and financial feedback.

The paper is structured as follows: Section 2 provides the theoretical background and introduces the research model; and Section 3 describes the data collection and methodological approach. The empirical results and their implications are presented in Sections 4 and 5, respectively, while Section 6 offers recommendations for research, policy, and system design. The paper concludes in Section 7.

# 2 Theoretical framework

## 2.1 Key theories connecting perception, actual behavior, and behavioral intentions

Motivations for using smart homes for energy management and behavior change can be better understood through the established behavioral theories. We conduct a comprehensive literature review to identify theories and models that help explain the relationships between user perception and intention to use Smart Home systems for energy efficiency (see Appendix A). The Theory of Planned Behavior (TPB) is the most widely used framework for describing the nexus between user perception and behavior. We use TPB along with the Norm Activation Model (NAM) to develop a theoretical framework that addresses the key factors influencing SHT usage.

The TPB, introduced by Ajzen in 1985 as an extension of Fishbein and Ajzen's earlier Theory of Reasoned Action (Ajzen, 1985; Fishbein and Ajzen, 1975), is a foundational model for predicting individual behavior. TPB posits that behavior is directly influenced by behavioral intentions, which are shaped by attitudes toward the behavior, subjective norms, and perceived behavioral control. Behavioral intention reflects an individual's willingness to carry out a specific behavior. The TPB framework has been widely applied in many fields, including energy conservation studies, where it helps explain intentions and behaviors related to saving energy (Greaves et al., 2013; Kim & Hwang, 2020). Its frequent use in information systems context further underscores its relevance and utility in understanding user interactions with technology (Anser et al., 2020; Arkorful et al., 2020; Chen, 2020), making TPB a robust theoretical foundation for investigating how individuals' intentions translate into specific actions.

NAM adds to the TPB by specifically addressing pro-environmental behaviors, making it useful for understanding actions aimed at sustainability (Onwezen et al., 2013; Schwartz, 1977). NAM focuses on the psychological factors behind altruistic and eco-friendly actions, identifying three main variables: awareness of the consequences, their ascription of responsibility for those consequences, and their personal norms that guide behavior (Han, 2014; Kim & Hwang, 2020). When combined with TPB, NAM enhances the analysis of environmental behaviors by adding depth to the understanding of how personal values and social responsibilities influence behavior (Kim & Hwang, 2020; Schwartz, 1977). This combined approach has proven effective in explaining not only general pro-social intentions but also specific energy-saving actions in various contexts (Chen, 2016). The application of NAM has been further refined by suggesting its use as a mediator model, where awareness of environmental consequences leads individuals to feel a sense of responsibility, thereby motivating eco-friendly behaviors (Groot & Steg, 2009). We build on this mediator model to provide a nuanced understanding of how increased environmental awareness can directly influence users' ascription of responsibility towards energy conservation.

## 2.2 Research model

In this section, we present a research model based on TPB and NAM (Figure 1). The model examines the factors that influence the **behavioral intention (BI)** to use Smart Home systems for energy efficiency. We present the research hypotheses in Table 1.

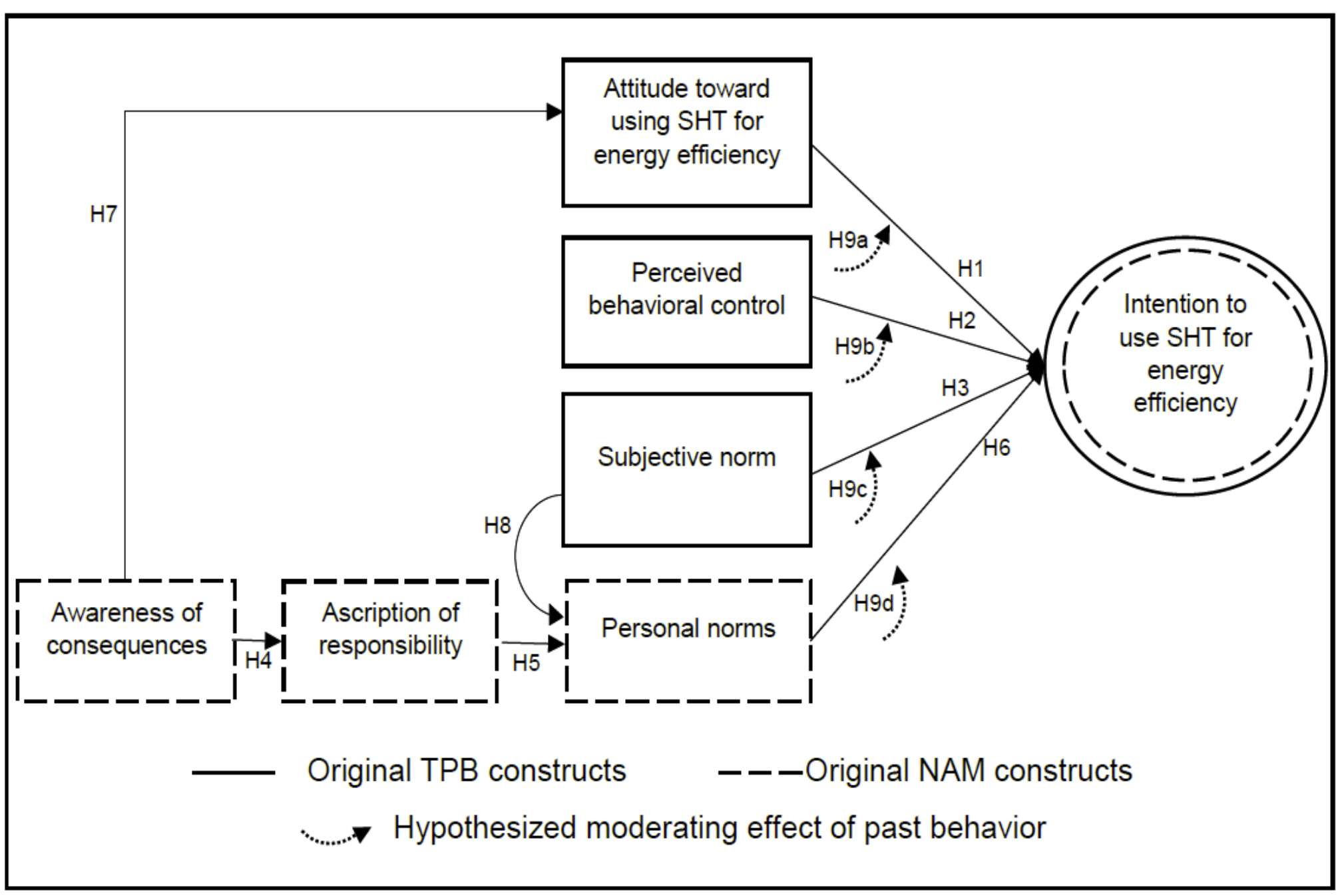

**Figure 1.** Proposed research model.

**Attitude (AT)** The user's *attitude* represents a person's evaluation of a particular behavior and the feelings they have when performing the behavior (Ajzen, 1985). It includes the consideration of potential benefits that the person believes are linked to the behavior (Greaves et al., 2013). A positive *attitude* towards the environment has often been shown to significantly affect energy-saving behaviors (Kim & Hwang, 2020; Wang et al., 2014). Thus, we hypothesize that users who hold positive attitudes toward using smart homes for energy efficiency are more inclined to engage in this kind of behavior (see Table 1).

**Perceived Behavioral Control (PBC)** *Perceived behavioral control* is defined as the individual's perception of their ability to perform a behavior successfully and is similar to the concept of self-efficacy (Ajzen, 1991). A user's control beliefs shape their *perceived behavioral control*. These control beliefs include a user's perceived ability and perceived difficulty of performing the behavior (Ajzen, 1991; Wang et al., 2014). Factors such as convenience, skills, time, mobility, and economic situation can influence energy-saving decisions (Anker-Nilssen, 2003). Thus, higher levels of *perceived behavioral control* regarding energy-saving behavior are associated with higher willingness to save energy (Oikonomou et al., 2009).

**Subjective Norm (SN)** *Subjective norm* describes the influence or social pressure that individuals perceive from their immediate environment, especially close family members or friends, regarding a behavior (Ajzen, 1985). Social norms, as well as personal ethical standpoints, are found to affect residents' energy-saving behavior (Wang et al., 2014; Webb et al., 2013). Hence, users who believe that people close to them support energy-saving behaviors are more likely to intend to engage in such behaviors.

**Awareness of Consequences (AC)** *Awareness of consequences* refers to an individual's perception of how their behavior affects others and the environment (Schwartz, 1977). People are more likely to act pro-environmentally if they understand the impact of their actions (Han & Hyun, 2017). Research shows that a person needs to be aware of the consequences of a behavior to develop feelings of responsibility and to engage in this behavior (Steg & Groot, 2010). Therefore, we hypothesize that users who are aware of the consequences of their energy consumption behavior develop more responsibility regarding the potentially positive consequences of acting pro-environmentally.

**Ascription of Responsibility (AR)** *Ascription of responsibility* refers to someone's belief that they are personally responsible for the outcomes of their behavior (Schwartz, 1977). People who see themselves as responsible for environmentally-related issues are more likely to support pro-environmental behaviors and engage in them (Rezaei et al., 2019; Shin et al., 2018). *Ascription of responsibility* mediates the relationship between *awareness of consequences* and *personal norms* (Groot & Steg, 2009). Thus, we hypothesize that feelings of responsibility regarding energy consumption behavior activate personal norms.

**Personal Norms (PN)** *Personal norms* refer to an individual's internal sense of moral obligation to act in a certain way (Schwartz, 1977). The moral obligation can act as a motivation to engage in pro-environmental behaviors (Onwezen et al., 2013). Research shows that *personal norms* influence pro-environmental behaviors (Kim & Hwang, 2020; Park & Ha, 2014; Zhang et al., 2017). *Personal norms* further mediate the relationship between *ascription of responsibility* and *behavioral intention* (Groot & Steg, 2009). We hypothesize that these *personal norms* regarding energy consumption behavior influence the b*ehavioral intention* to use smart homes for energy efficiency.

**Integrating TPB and NAM** Combining TPB and NAM provides deeper insights into eco-friendly *behavioral intentions* (Han & Hyun, 2017; Kim & Hwang, 2020; Onwezen et al., 2013). *Personal norms* can influence eco-friendly *behavioral intentions*, as well as *attitude*, *perceived behavioral control*, and *subjective norm* (Chen, 2016; Liu et al., 2017; Park & Ha, 2014). *Awareness of consequences* and *attitude* are shown to have a significant and positive association with each other (Kim & Hwang, 2020; Meng et al., 2020). In the context of smart homes, users should tend to have more positive attitudes towards using Smart Homes for energy efficiency if they are aware of the environmental consequences of higher energy consumption. Furthermore, a causal link between *subjective norms* and *personal norms* is supported by previous studies (Liu et al., 2017). It is argued that *subjective norms* precede *personal norms* since the social annotations of a specific behavior shape the way a person develops their belief about that behavior (Kim & Hwang, 2020; Onwezen et al., 2013).

**Past Behavior (PB)** *Past behavior* refers to actions taken in previous similar situations (Sommer, 2011). Previous studies include *past behavior* in TPB frameworks as a predictor of *behavioral intention* and actual behavior (Earle et al., 2020; Kidwell & Jewell, 2008). In some cases, *past behavior* is the strongest predictor of behavioral intention (Conner et al., 2003; Forward, 2009). *Past behavior* can also predict unique variance in intentions that is not linked to other TPB variables (Ouellette & Wood, 1998). For some studies, *past behavior* may be a better predictor for future behavior than intention (Sheeran, 2002; Wong & Mullan, 2009), which is based on the assumption that human behavior is driven by automatic processes rather than intentions (Sommer, 2011).

In the context of residential energy consumption, lifestyle and habit are two further important factors (Wang et al., 2014). Hence, the assumption is that users who have already behaved a certain way in the past, are more likely to continue behaving this way. Furthermore, feedback can affect behavioral, normative, and control beliefs which may influence future intentions about the underlying behavior (Ajzen, 2020). Thus, we hypothesize that *past behavior* moderates the strength of the relationships between the main variables and *behavioral intention*.

**Table 1.** Research Hypotheses

| Hypothesis | Variable | Theory |
|---|---|---|
| **H1:** *Attitudes* toward using Smart Home for energy efficiency positively influence users' *behavioral intentions*. | *Attitude (AT)* | TPB |
| **H2:** *Perceived behavioral control* positively influences users' *behavioral intentions* to use Smart Home for energy efficiency. | *Perceived behavioral control (PBC)* | TPB |
| **H3:** *Subjective norms* positively influence users' *behavioral intentions* to use Smart Home for energy efficiency. | *Subjective norm (SN)* | TPB |
| **H4:** *Awareness of consequences* positively influences *ascription of responsibility*. | *Awareness of consequences (AC)* | NAM |
| **H5:** *Ascription of responsibility* positively influences *personal norms* | *Ascription of responsibility (AR)* | NAM |
| **H6:** *Personal norms* positively influence the *behavioral intentions* to use Smart Home for energy efficiency. | *Personal norms (PN)* | NAM |
| **H7:** *Awareness of consequences* positively influences *attitudes* toward using Smart Home for energy efficiency. | *Integrating TPB and NAM* | combination of TPB and NAM |
| **H8:** *Subjective norms* positively influence *personal norms.* | *Integrating TPB and NAM* | combination of TPB and NAM |
| **H9a:** *Past behavior* moderates the effect of *attitude* on *behavioral intention*. | *Past behavior (PB)* | - |
| **H9b:** *Past behavior* moderates the effect of *perceived behavioral control* on *behavioral intention*. | *Past behavior (PB)* | - |
| **H9c:** *Past behavior* moderates the effect of *subjective norms* on *behavioral intention*. | *Past behavior (PB)* | - |
| **H9d:** *Past behavior* moderates the effect of *personal norms* on *behavioral intention*. | *Past behavior (PB)* | - |

# 3 Methods

## 3.1 Data collection and survey design

We develop a questionnaire based on the proposed research model and the findings from the literature review (see Appendix B). The online survey was conducted using LimeSurvey software. The target group were people who currently use Smart Home technology in their homes. Participants were recruited by sharing the survey link in online forums focused on topics around Smart Home (Facebook, Reddit, and various smart home forums). The questionnaire was made available in English and German to reach as many relevant respondents as possible within the scope of this work. Data collection took place between September and December 2021. A total of 466 people completed the survey (363 valid answers).

Survey questions were designed to reflect each hypothesis of the research model. Multiple items were created for each hypothesis to accurately capture the users' perceptions while keeping the questionnaire length reasonable. For most of the questions, response options were provided, and several questions offered the option for a complementary text entry.

Additionally, some questions aimed to collect information regarding the Smart Home systems in use and their functionalities, as well as questions regarding potential changes in users' total energy consumption.

To ensure participants were actual Smart Home users, the first question of the survey asked whether the participant's home was equipped with a Smart Home system. A seven-point Likert scale was used to measure the latent variables: *attitude*, *perceived behavioral control*, *subjective norm*, *personal norms*, *awareness of consequences*, and *ascription of responsibility*. *Past behavior* was measured using an eight-point Likert scale.

The items measuring *past behavior* were designed to reflect specific, relatable actions, as more concrete behavioral references tend to yield more accurate data than broad explanations (Oskamp & Schultz, 2005). The environmentally friendly behaviors for the questionnaire were chosen by considering which actual or desired behavioral changes were mentioned in the literature review (McIlvennie et al., 2020). Socio-demographic data was collected at the end of the survey.

All participants who either do not have a Smart Home system or could not specify one are excluded from the analysis. Furthermore, only responses with all answers were considered for the analysis. We also examined whether participants showed straight-lining behavior across items within the latent variable scales to identify potential careless responses. However, no notable patterns of such behavior were found. For questions offering a free text field, responses that did not fit any of the provided response options were merged with the most appropriate existing category. For example, one of the entries stated "disabled" as their current occupation status, which was not covered by the response options. Hence, this entry was grouped under "unemployed." The same process was applied for the socio-economic questions regarding education and home ownership.

Reverse scoring was applied for the fifth item of the latent variable *perceived behavioral control* due to the nature of the question. A high score on the Likert scale for this specific item would indicate lower *perceived behavioral control*, whereas a high score for the other items of this construct indicates higher *perceived behavioral control*, hence the reverse scoring. Finally, the data for the latent variables show a multivariate non-normal distribution.

### 3.2 Hybrid approach of structural equation modeling and random forest analysis

Structural equation modeling (SEM) is a common technique used in behavioral research to identify causal relationships between latent or measured variables (Hair et al., 2010). It has been applied in various contexts, including psychology, environment, and information systems (Ramlall, 2017; Ji & Chan, 2019; Pal et al., 2018; Park et al., 2017; Rastegari Kopaei et al., 2021). SEM is a framework that combines multiple statistical methods and can evaluate several dependence relationships simultaneously (Hair et al., 2010). It provides explanatory power by showing how well a theoretical model fits the underlying survey data and consists of a structural model and a measurement model. The former shows the causal relationships between endogenous and exogenous variables, while the latter reflects the relationships between latent factor variables and observed variables (Hair et al., 2021).

Attitudes and perceptions are main drivers of user intention, but these are latent constructs that cannot be directly measured. Hence, they are inferred through specific questionnaire items. SEM is particularly suited for analyzing latent constructs and offers advantages in addressing measurement errors compared to methods such as regression or factor analysis (Ramlall, 2017). Moreover, SEM is a flexible and comprehensive approach for testing theoretical models. Among the two dominant SEM approaches - covariance-based SEM (CB-SEM) and partial least squares SEM (PLS-SEM) - PLS-SEM is more suitable for smaller sample sizes, complex models, and non-normal data distributions (Hair et al., 2021). Additionally, PLS-SEM allows for the inclusion of advanced model components, which is beneficial for our research model. Therefore, we conduct the data analysis using PLS-SEM.

To build an explanatory model that can uncover both linear and non-linear causal relationships between the constructs in our proposed research model, SEM is extended with a machine learning technique. This method has been applied in several domains (Hizam et al., 2021; Priyadarshinee et al., 2017; Scott & Walczak, 2009; Sternad Zabukovšek et al., 2019). The combination of SEM and machine learning algorithms is often implemented as a two-step process during data analysis (Chong, 2013; Sternad Zabukovšek et al., 2019). SEM is used to test the hypothesized influence of independent variables on the dependent variable, while the machine learning part evaluates the relative strength of these effects (Leong, et al., 2020a).

Standard machine learning algorithms are typically not suited for causal analysis due to their black-box nature (Chan & Chong, 2012; Chong, 2013). To support causal interpretation, models must be transparent (Li et al., 2021). However, machine learning methods are particularly effective in detecting non-linear relationships between variables (Joshi & Yadav, 2018; Leong, et al., 2020b), while SEM is limited to linear relationships, which can oversimplify the factors influencing behavioral intention (Sohaib et al., 2020). Thus, machine learning helps uncover non-linear patterns within the proposed research model.

Several machine learning methods can be used in survey research (Buskirk et al., 2018; Kern et al., 2019)). Prior studies have integrated artificial neural networks with SEM (Priyadarshinee et al., 2017), as well as other techniques such as logistic regression, Bayesian networks, random forests, decision trees, if-then-else statements, and association rule mining (Akour et al., 2021; Arpaci, 2019; Dishakjian et al., 2021; Wong et al., 2021). In this study, we use a random forest algorithm to extend the SEM analysis, as it provides better interpretability and transparency in identifying the key factors that influence behavioral intention (Li et al., 2021). The random forest algorithm builds an ensemble of decision trees to predict an outcome based on multiple input variables (Breiman, 2001). The combined approach of SEM and random forest has been successfully applied in previous research to explain relationships between independent and dependent variables (Dishakjian et al., 2021; Shi et al., 2018).

### 3.2.1 Model specification

**Structural equation modeling** We use the SEMinR package (Ray et al., 2022) to estimate the SEM using the PLS-SEM method. The analysis is conducted in two steps: evaluating the measurement model and the structural model. First, the measurement model is defined by specifying the latent variables as constructs and describing their indicators which are derived from the questionnaire items. The latent variables *awareness of consequences*, *ascription of responsibility*, and *personal norm*s are each measured using three indicators. *Attitude* is measured by seven indicators, *subjective norm* by five, *perceived behavioral control* by six, *behavioral intention* by five, and *past behavior* by four. We assume a reflective measurement, meaning the indicators are effects of the underlying construct, as opposed to the construct being an outcome of the indicators.

Second, the structural model is defined, which outlines the relationships and directional paths between constructs. The method computes partial regression models for each path based on the survey data, producing correlations between constructs and their indicators. Due to the nonparametric characteristic of the PLS-SEM method, we apply bootstrapping to estimate standard errors and compute confidence intervals. This is done by generating multiple subsamples from the original dataset to estimate the model parameters.

Third, we evaluate the measurement model for reliability and validity. In particular, we assess the indicator reliability, which explains how well each survey item (indicator) reflects its underlying construct. This is measured using indicator loadings - the correlation between the indicator and the construct it is intended to measure. Indicator loadings above 0.708 indicate good reliability, as they explain more than 50% of the indicator's variance (Hair et al., 2021). Internal consistency reliability is then checked using composite reliability, Cronbach's alpha, and the reliability coefficient rhoA, which measure whether indicators of the same construct are correlated. Convergent validity assesses whether a construct sufficiently explains the variance in its indicators and is evaluated using average variance extracted (AVE). Values above 0.5 are considered acceptable (Hair et al., 2019). Discriminant validity, which examines how distinct constructs are from each other, is assessed using the heterotrait-monotrait ratio (HTMT) of correlations. The HTMT compares the average correlations across constructs with the average correlations within the same construct. Values should not exceed 0.9 for conceptually similar constructs (Henseler et al., 2015).

Fourth, we assess the structural model quality by evaluating collinearity, path significance, explanatory power, and predictive accuracy. Collinearity examines whether constructs are highly correlated, which can distort interpretation. The variance inflation factor (VIF) identifies collinearity issues when values exceed 5 (Hair et al., 2021). The path coefficients are evaluated for statistical significance (t-value) and effect size ($f^2$). Bootstrapping is used to calculate t-values and confidence intervals for each coefficient. Explanatory power is assessed using the coefficient of determination ($R^2$), which indicates how much variance is explained in the endogenous constructs (Shmueli & Koppius, 2011). Higher $R^2$ values indicate better explanatory strength. Predictive power is assessed using k-fold cross-validation, where the data is randomly split into subsets and repeatedly tested. Common prediction error metrics include root mean square error (RMSE) and mean absolute error (MAE).

Lastly, since our model includes the moderator variable *past behavior*, we also conduct a moderation analysis. This requires introducing interaction terms, as a moderator not only affects the endogenous construct directly but also changes the relationship between an exogenous and endogenous variable (Hair et al., 2021). We include four interaction terms for the relationships between *attitude, subjective norm, perceived behavioral control*, and *personal norms* with *behavioral intention*.

**Random Forest analysis**

We use the significant variables from the SEM results to conduct a random forest analysis and explore non-linear relationships with *behavioral intention*. We conduct the analysis in R using the randomForest package (Liaw & Wiener, 2002), which implements Breiman's original algorithm. The caret package (Kuhn, 2024) is used to tune and optimize model parameters.

The indicators from the SEM model are used as predictors. The average score of the four *behavioral intention* variables (BI1, B2, B3, and BI4) is used as the outcome variable, ranging from 1 (low intention) to 7 (high intention) to use Smart Home for energy efficiency, making the random forest a regression model. From the preprocessed dataset, 70% of the responses are used for training and 30% for validation, resulting in 254 training samples and 109 validation samples. A bagging approach is used, which builds multiple trees from bootstrap samples of the training data. This helps reduce variance by averaging multiple noisy but unbiased models (Hastie et al., 2009). For each tree, a random subset of predictors is chosen at each split. To prevent overfitting, we apply ten-fold cross-validation. After the model is trained, variable importance is calculated by measuring the improvement in the splitting criterion at each node. The average importance across all trees provides a ranking of the predictors.

To optimize model performance, we tune two main parameters: the number of predictors considered at each split (*mtry*) and the minimum number of samples required at a terminal node (nodesize). A grid search is used to test *mtry* values from 1 to 8 and nodesize values of 1, 3, 5, and 10. Another key parameter is the number of trees. Increasing the number of trees beyond 750 does not improve performance (for details see Figure D1 in Appendix). Thus, the number of trees is set to 750 to reduce training time. To assess model accuracy, we report RMSE, $R^2$, and MAE as performance metrics. The programming code used in this work (and beyond) is open-sourced [Link to GitHub will be added after the peer-review process].

# 4 Results

## 4.1 Key demographics and descriptive statistics of respondents

Among 363 observations, most of the survey participants are male (94%). The age of the participants is approximately normally distributed with the biggest group within the middle-aged category of 35-44 years (30%), followed by the age groups 45-54 (26%), and 25-34 (20%). In terms of educational background, 37% have a high school diploma or less, 36% hold a bachelor's degree, and 22% received a master's degree. Regarding the occupation status, 84% of all survey participants are either employed full-time or self-employed, whereas the remaining participants are either retired (9%), or students, unemployed, or part-time employed (6%). The majority of survey participants are homeowners (82%). Geographically, the distribution reflects the language options that were provided in the survey. Around 53% of all participants are from Germany and 30% are people from English-speaking countries (USA, Canada, UK, Australia). Household sizes consist mainly between two people (40%) and 3-4 people (42%). Most of the households do not have any children (61%), followed by those with one or two children (35%). Regarding the monthly net income of the household, 26% fall in the range of 5K-8K USD, followed by 22% within 3K-5K USD. A substantial number of participants report higher monthly income of 8K USD and more (see Table C1 in Appendix for details).

Since SH systems can vary greatly between households due to its modular nature, additional information about the participants' SH systems have been gathered. Most of the participants have been using SH technology between one and three years (43%), followed by four to six years (21%) and seven to ten years (11%). Recent users who have been using their SH for less than a year make up around 17%, whereas 6% have only started less than six months ago. The SH systems Home Assistant, Homematic IP, OpenHAB, HomeKit, ioBroker, and KNX, as well as products from Amazon, Google, and Philips were most often mentioned. Many participants also mentioned that they are using a combination of several systems. Out of all participants, around 59% state that they are currently not using a SHEMS as part of their smart home, whereas 39% state that they do. The most popular sensors and IoT products installed in the participants' homes include smart plugs, light sources, switches, motion sensors, and smart thermostats. We provide a complete list of the smart home products and their respective occurrences across the sample size in Table C2 in Appendix.

Furthermore, participants were asked what motivates them to install a SH (Table C3 in Appendix). Participants had to rank ten different reasons by allocating a total of 100 points. The most important factor is the interest in SH technology, closely followed by the increase in comfort and convenience. The sustainability and energy saving, entertainment-related reasons, desire to have more control over the home and to increase safety and security are ranked in third to sixth place respectively. Monetary savings are only ranked in seventh place, whereas homecare (ambient-assisted living) plays only a minor role. The "other" option includes the desire to install and program the system by residents, recommendation by friends, or other convenience- and interest-related motivations.

## 4.2 Structural model results: personal norms, responsibility, and awareness drive energy-saving intentions

**Assessment of Measurement Model** First, the measurement model without the moderator *Past Behavior* is analyzed to focus on the main effects on *Behavioral Intention*. The initial measurement model included all items from the questionnaire. However, after assessing the different measures, some of them showed low indicator loadings. For social science studies, indicator loadings below the threshold are not uncommon and should only be excluded if this improves other assessment factors (Hair et al., 2021). As this is the case, the following indicators are removed: *Attitude:* AT1, AT2, AT3, AT6; *Subjective Norm:* SN4, SN5; *Perceived Behavioral Control:* PBC2, PBC3, PBC5, PBC6; *Behavioral Intention:* BI5. The results show that the exclusion of these indicators below the threshold of 0.708 yield better results for internal consistency reliability, convergent validity, and discriminant validity. Moreover, the remaining indicators offer sufficient content validity to represent the respective constructs. The exclusion of these indicators implies that the corresponding items in the questionnaire have not been able to properly capture the desired constructs and may have to be revised if used in future research.

The indicator loadings for all items are now above the recommended threshold of 0.708 which provides acceptable indicator reliability (Hair et al., 2021), see Table 1. Cronbach's alpha and rhoA show overall good results for the internal consistency reliability with values between 0.7 and 0.9. Regarding the convergent validity, all variables display average variance extracted (AVE) values greater than 0.5 which is the minimum acceptable AVE. For discriminant validity, we analyze heterotrait-monotrait ratio (HTMT) of correlations values (Table 2). Conceptually similar constructs with values over 0.9 indicate problems in discriminant validity. This is the case for the HTMT value for *Awareness of Consequences* and *Ascription of Responsibility* (0.92), and *Ascription of Responsibility* and *Personal Norms* (1.02). These constructs all stem from the NAM and have substantial similarities. Especially due to the underlying research question, the differentiation of the questions that are supposed to capture each of these constructs are vague. However, these constructs and their corresponding items will stay in the measurement model since they are expected to be highly correlated, measuring theoretically similar concepts.

**Table 1.** Reliability and convergent validity of measurement model.

| Latent Variable | Measurement Indicator | Loading (≥0.708) | Cronbach's Alpha (0.7~0.9) | Reliability Coefficient ($rho_A$) (0.7~0.9) | AVE (≥0.5) |
|---|---|---|---|---|---|
| Attitude | AT4 | 0.85 | 0.84 | 0.84 | 0.76 |
| | AT5 | 0.89 | | | |
| | AT7 | 0.87 | | | |
| Subjective Norm | SN1 | 0.86 | 0.79 | 0.81 | 0.70 |
| | SN2 | 0.81 | | | |
| | SN3 | 0.84 | | | |
| Perceived Behavioral Control | PBC1 | 0.92 | 0.70 | 0.75 | 0.77 |
| | PBC4 | 0.83 | | | |
| Awareness of Consequences | AC1 | 0.84 | 0.83 | 0.84 | 0.75 |
| | AC2 | 0.91 | | | |
| | AC3 | 0.85 | | | |
| Ascription of Responsibility | AR1 | 0.85 | 0.80 | 0.81 | 0.72 |
| | AR2 | 0.90 | | | |
| | AR3 | 0.78 | | | |
| Personal Norms | PN1 | 0.80 | 0.77 | 0.77 | 0.69 |
| | PN2 | 0.81 | | | |
| | PN3 | 0.87 | | | |
| Behavioral Intention | BI1 | 0.82 | 0.87 | 0.88 | 0.72 |
| | BI2 | 0.91 | | | |
| | BI3 | 0.87 | | | |
| | BI4 | 0.80 | | | |

**Table 2.** Discriminant validity of measurement model.

| | Awareness of Consequences | Ascription of Responsibility | Subjective Norm | Attitude | Perceived Behavioral Control | Personal Norms |
|---|---|---|---|---|---|---|
| Ascription of Responsibility | **0.92** | | | | | |
| Subjective Norm | 0.78 | 0.78 | | | | |
| Attitude | 0.45 | 0.59 | 0.47 | | | |
| Perceived Behavioral Control | 0.19 | 0.23 | 0.20 | 0.48 | | |
| Personal Norms | 0.83 | **1.02** | 0.80 | 0.62 | 0.27 | |
| Behavioral Intention | 0.75 | 0.86 | 0.73 | 0.73 | 0.38 | 0.89 |

**Assessment of Structural Model** After confirming the reliability and validity of the measurement model, we inspect the collinearity of the structural model. All variance inflation factor (VIF) values are below the threshold of three. This suggests that there are no collinearity problems between the constructs that influence endogenous variables. For the relationships between the latent variables, the original path coefficient estimates indicate the strongest positive influence from *Awareness of Consequences* to *Ascription of Responsibility* (0.76) and the lowest positive influence from *Perceived Behavioral Control* to *Behavioral Intention (0.07)*, see Table 3. Bootstrapping the structural paths and assuming a significance level of five percent, the t-values reveal statistical significance for all variable relationships. However, the relationship between *Perceived Behavioral Control* and *Behavioral Intention* presents a very low path coefficient with a t-value and p-value barely above the threshold for statistical significance. This is emphasized by the low effect size ($f^2$=0.01) for *Perceived Behavioral Control* on *Behavioral Intention*. Thus, *Perceived Behavioral Control* is assumed to only have a minimal effect on *Behavioral Intention*. The values for the coefficient of determination ($R^2$) of the endogenous latent variables, which assess the model's explanatory power, present moderate effects for *Ascription of Responsibility* (0.58), *Personal Norms* (0.67), and *Behavioral Intention* (0.65), whereas a weak effect is implied for *Attitude* (0.15). Comparing RMSE values to analyze the predictive power of the model, only one of the indicators displayed a lower RMSE compared to the naïve linear regression model benchmark. Therefore, the model has low predictive power. Additionally, the mediation effect from *Subjective Norm* on *Personal Norms* and *Behavioral Intention* is analyzed by testing the significance of indirect effects in the model. The results show that each path of the mediation effect is statistically significant, and the product of all paths is positive. This suggests a complementary mediation effect which means that the indirect and direct effect point in the same direction.

**Table 3.** Assessment of structural model.

| Hypothesis | Relationship | Path coefficient | Bootstrap SD | t-value ($>\lvert 1.96 \rvert$) | p-value | Supported | $R^2$ |
|---|---|---|---|---|---|---|---|
| H1 | AT →BI | 0.31 | 0.05 | 6.55 | < .001 | Yes | 0.65 |
| H2 | PBC →BI | 0.07 | 0.03 | 2.00 | 0.02 | Yes | |
| H3 | SN →BI | 0.21 | 0.04 | 5.00 | < .001 | Yes | |
| H6 | PN →BI | 0.42 | 0.05 | 9.17 | < .001 | Yes | |
| H4 | AC →AR | 0.76 | 0.03 | 27.48 | < .001 | Yes | 0.58 |
| H5 | AR →PN | 0.66 | 0.03 | 18.59 | < .001 | Yes | 0.67 |
| H8 | SN →PN | 0.22 | 0.04 | 5.40 | < .001 | Yes | |
| H7 | AC →AT | 0.38 | 0.06 | 6.16 | < .001 | Yes | 0.15 |

Note: Bootstrap sample=10000. Significance level α=5%. Abbreviations: BI - Behavioral Intention, AT - Attitude, PBC - Perceived Behavioral Control, SN - Subjective Norm, AC - Awareness of Consequences, AR - Ascription of Responsibility, PN - Personal Norms, PB - Past Behavior.

**Moderation analysis** The moderating effect of *Past Behavior* on the constructs *Attitude*, *Subjective Norm*, *Perceived Behavioral Control*, and *Personal Norms* is investigated by first introducing interaction terms for each combination. Subsequently, the measurement and structural model are evaluated based on the common assessment criteria. For the measurement model, the exclusion of the aforementioned indicators, as well as PB2 and PB4 (*Past Behavior*) yield the best results in terms of reliability and validity. Apart from that, the measurement model shows no big differences compared to the measurement model without the moderator. Similar to the structural model without moderation, the moderated structural model shows no collinearity problems and moderate explanatory power for the endogenous variables *Ascription of Responsibility*, *Personal Norms*, and *Behavioral Intention*, and weak explanatory power for *Attitude*. The predictive power of the model is low, as only two of the indicators for *Behavioral Intention* (BI1 & BI3) have lower RMSE values than the benchmark of the linear model. The interaction term *Personal Norms* * *Past Behavior* indicates a large effect size with a value of 0.025 which explains how much the moderator adds to the effect of the endogenous construct. The other interaction terms do not show any meaningful effect sizes. Bootstrapping the model shows that the interaction term *Personal Norms* * *Past Behavior* has a small negative effect on *Behavioral Intention* (-0.13), whereas the interaction terms *Attitude* * *Past Behavior* (0.02), *Subjective Norm* * *Past Behavior* (0.03), and *Perceived Behavioral Control* * *Past Behavior* (0.02), as well as the construct *Past Behavior* (0.13), have positive effects on *Behavioral Intention*, albeit mainly minor (see Table 4). However, the t-values of the interaction terms show that only *Personal Norms* * *Past Behavior* is statistically significant, leading to the rejection of the other three hypotheses. Furthermore, this model also rejects the effect from *Perceived Behavioral Control* to *Behavioral Intention* which emphasizes the result of the minimal effect in the assessment of the previous structural model. Hence, *Past Behavior* moderates solely the relationship between *Personal Norms* and *Behavioral Intention*.

**Table 4.** Assessment of moderation effect.

| Hypothesis | Relationship | Path Coefficient | Bootstrap SD | T-Value (>\|1.96\|) | p-Value | Supported | $R^2$ |
|---|---|---|---|---|---|---|---|
| H4 | AC →AR | 0.76 | 0.03 | 27.62 | < .001 | Yes | 0.58 |
| H5 | AR →PN | 0.66 | 0.03 | 18.58 | < .001 | Yes | 0.67 |
| H8 | SN →PN | 0.22 | 0.04 | 5.40 | < .001 | Yes | |
| H7 | AC →AT | 0.38 | 0.06 | 6.22 | < .001 | Yes | 0.15 |
| H1 | AT →BI | 0.26 | 0.05 | 5.54 | < .001 | Yes | 0.68 |
| H2 | PBC →BI | 0.06 | 0.04 | 1.41 | 0.08 | No | |
| H3 | SN →BI | 0.21 | 0.04 | 5.00 | < .001 | Yes | |
| H6 | PN →BI | 0.38 | 0.04 | 9.24 | < .001 | Yes | |
| H9a | AT*PB →BI | 0.02 | 0.05 | 0.38 | 0.35 | No | |
| H9c | SN*PB →BI | 0.03 | 0.04 | 0.70 | 0.24 | No | |
| H9d | PN*PB →BI | -0.13 | 0.05 | -2.53 | 0.01 | Yes | |
| H9b | PBC*PB →BI | 0.02 | 0.03 | 0.67 | 0.25 | No | |
| | PB →BI | 0.13 | 0.04 | 3.24 | 0.001 | | |

Note: Bootstrap sample=10000. Significance level α=5%. BI - Behavioral Intention, AT - Attitude, PBC - Perceived Behavioral Control, SN - Subjective Norm, AC - Awareness of Consequences, AR - Ascription of Responsibility, PN - Personal Norms, PB - Past Behavior.

We show the final path model in Figure 2 with the green paths representing positive relationships between the constructs, whereas the red path represents a negative relationship. Constructs whose statistical significance could not be proven are grayed out. The results of the SEM analysis show that *Personal Norms* have the strongest effect on users' *Behavioral Intention*. *Personal Norms* are positively influenced by *Ascription of Responsibility* and, to a lesser extent, by *Subjective Norm*. *Awareness of Consequences* also shows a high impact on *Ascription of Responsibility*. However, since the measurement model initially showed a lack of discriminant validity for these relationships, the effects might be slightly overestimated. *Awareness of Consequences* also positively influences *Attitude*. People who have a positive perception about using their SH for energy efficiency show higher intentions to engage in this behavior. Having close friends and family that are supporting these behaviors has a positive effect on users' *Behavioral Intention* as well. The effect of *Perceived Behavioral Control* is not supported by the SEM. The moderating effect of *Past Behavior* could only be supported in the context of *Personal Norms*. The direct effect of *Past Behavior* on *Behavioral Intention* indicates a low positive influence. The interaction effect of *Past Behavior* and *Personal Norms*, however, shows a small negative influence on the relationship between *Personal Norms* and *Behavioral Intention*. These results suggest that the relationship between *Personal Norms* and *Behavioral Intention* is 0.38 if *Past Behavior* is moderate. For higher values of *Past Behavior*, the relationship between *Personal Norms* and *Behavioral Intention* decreases by the size of the interaction term (i.e., 0.38 - 0.13 = 0.25). Vice versa, lower values of *Past Behavior* result in an increase of the relationship between *Personal Norms* and *Behavioral Intention* by the size of the interaction term (i.e., 0.38 - (-0.13) = 0.51). Overall, the SEM analysis reveals moderate explanatory power for *Behavioral Intention*. Eight out of twelve hypotheses were shown to be statistically significant.

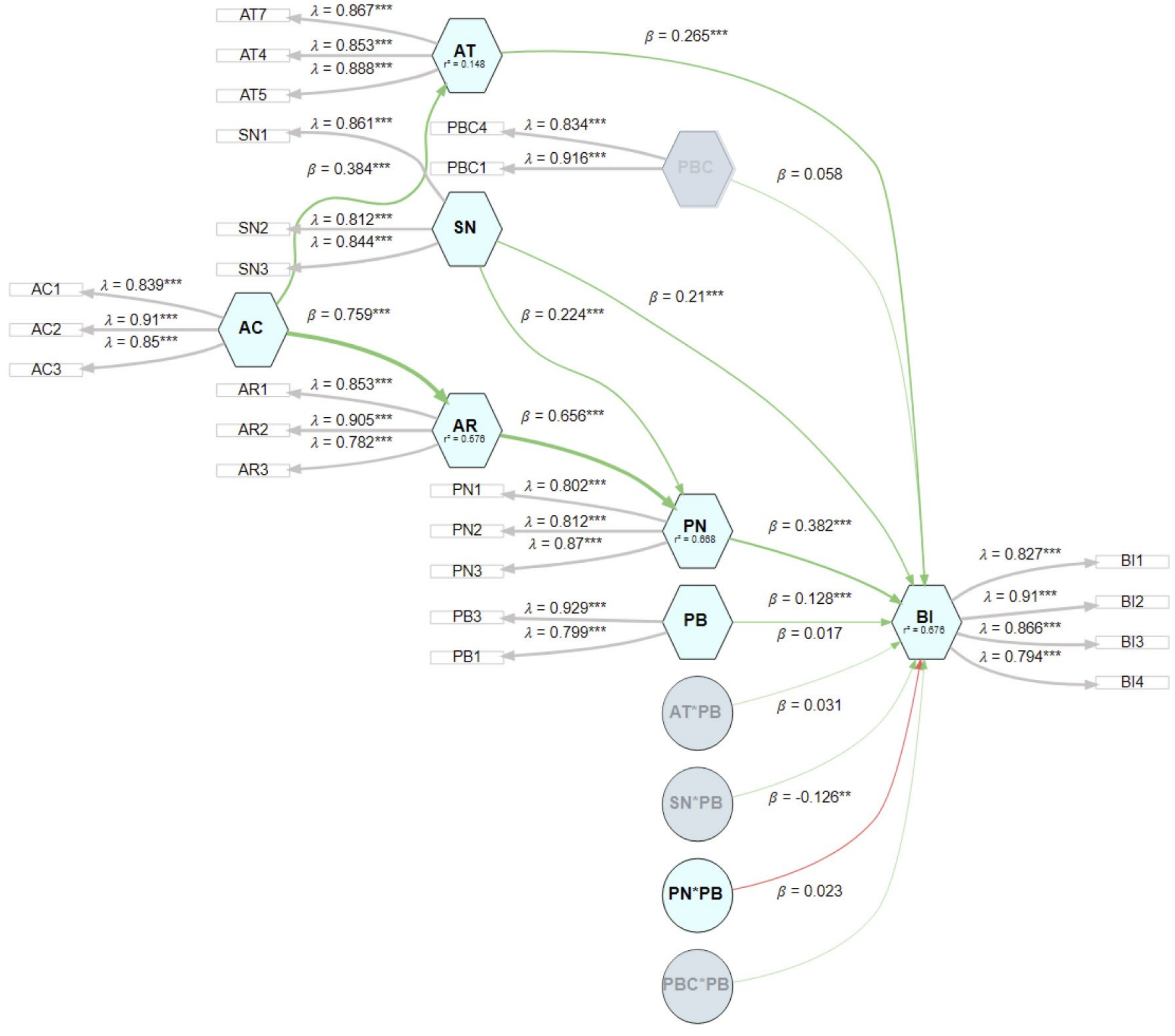

**Figure 2.** Path model with path coefficients and indicator loadings. BI - Behavioral Intention, AT - Attitude, PBC - Perceived Behavioral Control, SN - Subjective Norm, AC - Awareness of Consequences, AR - Ascription of Responsibility, PN - Personal Norms, PB - Past Behavior.

### 4.3 Random forest results confirm responsibility and personal norms as key predictors for smart home energy use intentions

As a second step, the SEM analysis is extended by applying a random forest algorithm to examine the importance of influential factors on *Behavioral Intention* to use SH for energy efficiency. A node size of three and an *mtry* of one yield the lowest RMSE value and are therefore selected as optimal parameters for the final model (Figure D2 in Appendix). The results of the final model show an RMSE of 0.5 and an $R^2$ value of 0.68 on the training data. For the validation set, the RMSE increases to 0.85, $R^2$ is 0.61, and the MAE is 0.65. As with the dependent variable *Behavioral Intention*, these error values are based on a seven-point Likert scale. This indicates that the model explains approximately 61% of the variance in the validation data, with an average prediction error (MAE) of 0.65. While the RMSE of around 0.8 suggests some prediction uncertainty - likely due to noise in the survey data - the model still demonstrates moderate predictive power in the context of human behavior and intention modeling.

Moreover, the model results help to understand the importance of the variables that affect user's intention to use their smart homes for energy efficiency. To assess this, we examine the variable importance from the trained model. The node purity values of the ten most important indicators are shown in Figure 3. Among them, AR2 (*Ascription of Responsibility*) has the highest importance, followed by PN3 and PN1 (*Personal Norms*). PN2 (*Personal Norms*), AT4 (*Attitude*), AC2 (*Awareness of Consequences*), and PB3 (*Past Behavior*) show similar levels of importance. The remaining top indicators include AR3 (*Ascription of Responsibility*), AT7 (*Attitude*), and SN1 (*Subjective Norm*).

Afterward, the indicators for each construct were aggregated and averaged to evaluate the overall importance of each construct (Figure 4, left). In addition, a second random forest model was trained using only the aggregated construct scores. In this case, the construct values were summed and averaged before model training. The results show a nearly identical ranking, except for *Subjective Norm* and *Past Behavior,* which switched places (Figure 4, right). Furthermore, the differences between the constructs are greater in this second model. Overall, *Ascription of Responsibility* ranks highest, followed by *Personal Norms*, *Attitude*, and *Awareness of Consequences*, reflecting similar patterns as observed in the SEM.

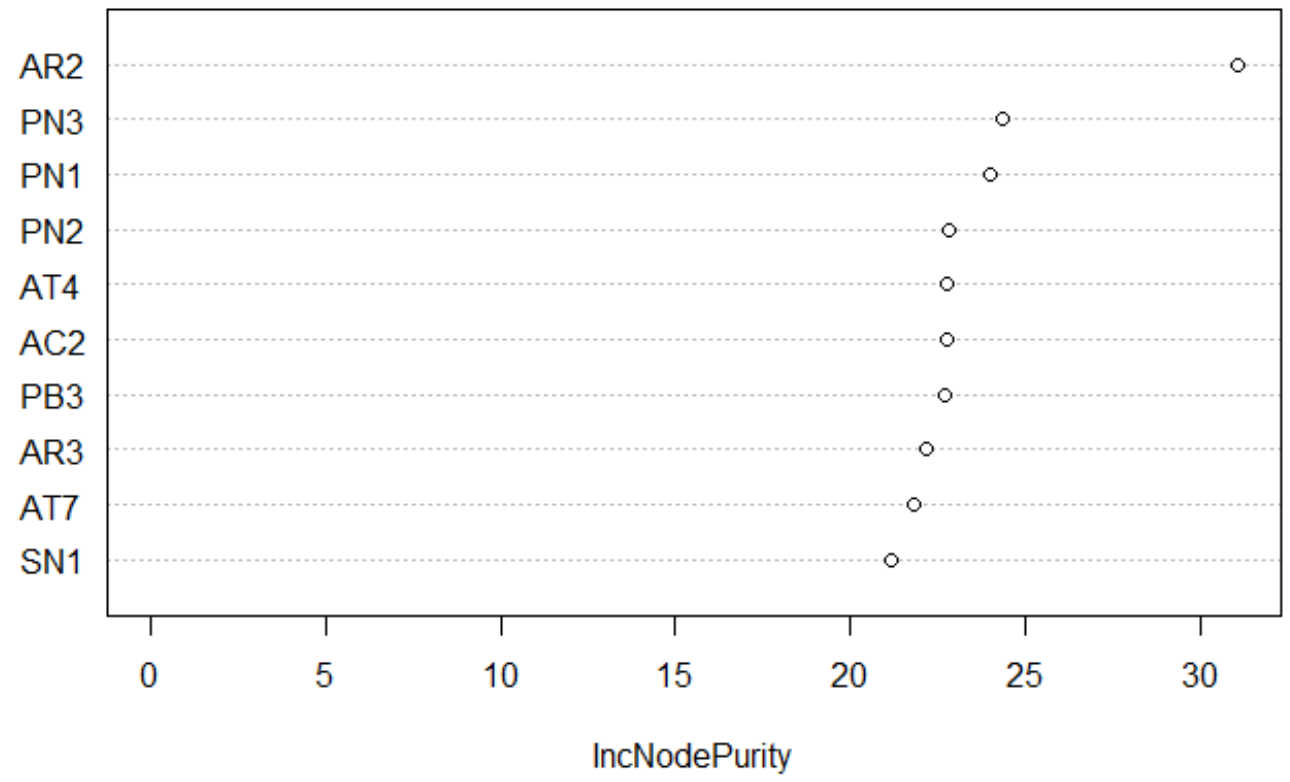

**Figure 3.** Indicator importance based on random forest results.

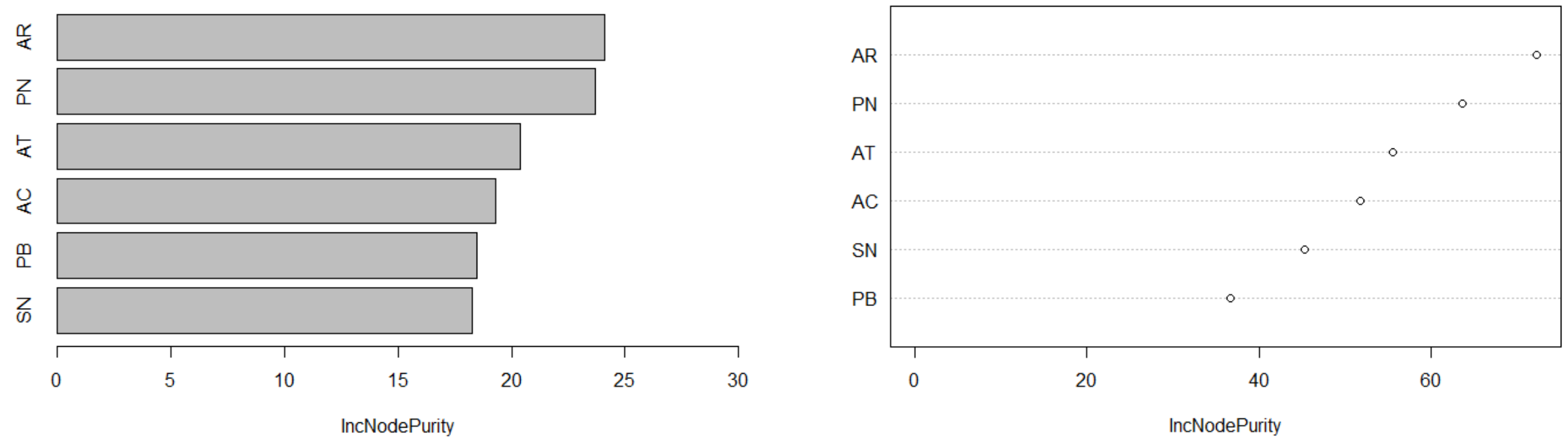

**Figure 4.** Left - Average variable importance based on individual indicators. Right - Variable importance based on newly trained random forest using aggregated construct scores. Abbreviations: BI - Behavioral Intention, AT - Attitude, PBC - Perceived Behavioral Control, SN - Subjective Norm, AC - Awareness of Consequences, AR - Ascription of Responsibility, PN - Personal Norms, PB - Past Behavior.

### 4.4 Survey insights: half report energy savings with smart home, many seek financial incentives

Survey participants were asked about their energy bills, energy consumption, and which smart home features they believe could encourage more energy-efficient behavior at home. About half (50.1%) of respondents reported noticing cost savings on their energy bills since installing smart home technology. By contrast, 26% said they had not noticed any savings, while the remaining 24% either did not observe savings or were not using smart home products likely to generate them. Around 49% of participants stated that their energy consumption had decreased since using a smart home. Approximately 7% reported an increase in consumption, and 27% observed no change. The remaining 18% said they did not know whether their energy use had changed.

Regarding incentives, participants indicated that the most motivating factors for reducing energy consumption would be receiving rewards (e.g., tax deductions) based on energy savings and getting information from their smart home about potential financial savings (Table C4 in Appendix). These were followed by clearer and more personalized guidance on how to save energy, comparisons with neighborhood energy consumption, and information about the environmental impact of reducing personal energy use. In contrast, a more gamified user interface and the ability to set personal energy goals were seen as the least motivating factors. As for the perceived usability of their smart home system (Table C5 in Appendix), the majority of participants rated it as "good" (42%), followed by "okay" (32%), and "very good" (20%). Only 6% rated the usability as "poor" or "very poor."

# 5 Discussion

## 5.1 Interpretation of results

The results of the SEM indicate that *Personal Norms* have the strongest influence on the intention to use smart homes for energy efficiency. Moreover, the positive effect of *Awareness of Consequences* on *Ascription of Responsibility*, followed by the effect of *Ascription of Responsibility* on *Personal Norms*, is statistically significant. This suggests that individuals who are more aware of the consequences of their energy consumption feel a stronger sense of responsibility, which in turn activates their *Personal Norms* and is reflected in their *Behavioral Intention* to use smart homes for energy efficiency. These findings align with previous research (Nilsson et al., 2018; Schwartz et al., 2015) highlight how energy feedback systems can increase awareness of energy consumption, thereby influencing users' potential to save energy.

The effects of *Subjective Norm* on *Behavioral Intention*, as well as on *Personal Norms*, were found to be statistically significant. This indicates that social pressure from close friends and family not only influences users' intention to use smart homes for energy efficiency but also shapes their moral obligation to behave more sustainably. These findings are consistent with Dominicis et al. (2019), who show that providing normative feedback (i.e., feedback framed as a social comparison) leads to long-term reductions in household electricity consumption. Similarly, Shevchuk et al. (2020) find that social support, when combined with perceived persuasiveness, positively influences the continued intention to use smart metering devices. In their study, perceived persuasiveness - defined as users' favorable impressions of smart the metering system - closely aligns with the construct of *Attitude* in this research, thereby supporting the observed positive effect of *Attitude* on *Behavioral Intention*.

These findings are further supported by the results of the random forest analysis. The variable importance ranking shows that *Ascription of Responsibility* scores highest, followed by *Personal Norms, Attitude*, and *Awareness of Consequences*. Since *Awareness of Consequences*, *Ascription of Responsibility*, and *Personal Norms* originate from the Norm Activation Model (NAM), the results suggest that pro-environmental sentiments are strong predictors of the intention to use smart homes for energy efficiency. This is reinforced by the item-level random forest results, which identify AR2 (*Ascription of Responsibility*) as the most important indicator, followed by PN3, PN1, and PN2 (*Personal Norms*). Each of these items emphasizes an individual's moral obligation to act sustainably. Thus, personal values appear to play the most crucial role in shaping the desire to adopt smart home technologies for environmental reasons.

The moderating effect of *Past Behavior* is only supported in the relationship between *Personal Norms* and *Behavioral Intention*, and it shows a negative influence. This suggests that when users have already established routines in their homes, *Personal Norms* have less impact on their intention to use smart homes for energy efficiency. In contrast, if users have not yet formed strong routines, the impact of their *Personal Norms* on *Behavioral Intention* increases. Similar findings are reported by Wang et al. (2014), who shows that residential habits are negatively correlated with the intention to engage in energy-saving behavior.

The lack of evidence for a relationship between *Perceived Behavioral Control* and *Behavioral Intention* may be explained by the relatively homogeneous sample of participants, who are likely early adopters of smart home technology and therefore tend to be more tech-savvy. As a result, participants may generally perceive a high level of control due to their technical skills. Early adopters are typically characterized by greater prior awareness of smart home technology, higher incomes, and active information-seeking behavior (Hargreaves & Wilson, 2017; Rogers, 2010). These characteristics align with the demographics of this study, which include a high proportion of homeowners and individuals with above-average incomes. Furthermore, interest in SHT was ranked as the most important reason for adoption. Schwartz et al. (2015) and Hargreaves et al. (2010) also highlight the predominance of male users in energy-related technology use, which is also the case for this study. Additionally, individuals with higher incomes who own their homes and are more aware of their energy costs may be more inclined to adopt energy-efficient behaviors (Karlin et al., 2014). Therefore, the observed effects in this study may not generalize to populations with different characteristics. It should also be noted that the reported household income data may be skewed due to inaccurate responses, as the values appear relatively high.

## 5.2 Implications

Based on the results, we draw three *implications*.

**(i) Pro-environmental intentions and behavior**

Strengthening public awareness of climate change and the role of household energy consumption can enhance pro-environmental intentions and behaviours. Framing smart home technologies as tools for energy efficiency - rather than mere convenience - can positively shape user motivation and behavioural intention. Informational strategies that highlight environmental impacts, combined with personalised energy feedback, may increase individuals' sense of responsibility and activate personal norms leading to energy-saving behaviors. Policy mechanisms such as financial incentives (e.g. tax credits or rebates tied to energy savings), along with targeted communication campaigns, could further support the adoption and effective use of smart homes for energy efficiency. Survey responses reinforce this potential, with a majority of participants indicating that cost savings and reward schemes would be strong motivators for engaging in energy-efficient behaviour via smart home systems.

**(iii) User-centric smart home design**

As end-users are central to achieving household energy savings, the design of smart homes must prioritise usability and adaptability. Systems should deliver information that is both easy to interpret and contextually relevant, enabling users to understand and act upon their consumption data. Survey responses support this need, with many participants expressing a preference for more accessible and personalised energy feedback. To support sustained engagement, smart home systems should account for diverse levels of technical skill, familiarity, and motivation, while also adapting to changes in user behavior over time. Tailored onboarding and support services can reduce adoption barriers, especially for users less experienced with digital technologies. Importantly, challenges such as poor interoperability across devices must be addressed to streamline installation and everyday use. Product preference data from the survey further highlight that users tend to favour affordable and easy-to-install devices over complex, high-cost alternatives. Lowering cognitive and technical barriers to adoption may therefore play a key role in unlocking the energy-saving potential of smart homes at scale.

**(iii) Systems-level approach**

Maximising the energy efficiency potential of smart homes requires a more holistic systems perspective. Outcomes are shaped not only by user behaviour but also by a complex interplay of structural, environmental, and socio-technical factors. Building characteristics (e.g., type, age, thermal performance) interact with local climate conditions, utility infrastructure, and occupant needs. The effectiveness of smart home technologies further depends on their integration into broader systems, including smart grids, utility platforms, dynamic tariff schemes, and demand response programmes. Policy and regulatory frameworks are critical enablers, particularly in addressing disparities linked to geography, housing stock, and socio-economic status. Ensuring equitable access and interoperability across systems is essential. Future research should consider these interdependencies to design inclusive, adaptive, and resilient smart home solutions that contribute meaningfully to energy system decarbonisation.

## 5.3 Recommendations

Drawing on the findings of this study, we propose a set of *recommendations* aimed at increasing the effectiveness and adoption of smart home systems for energy efficiency.

**(1) Leveraging personal norms rather than control beliefs**

The results suggest that personal norms, such as a sense of responsibility for reducing environmental impact, are stronger predictors of behavioral intention than perceived control. While technical usability remains essential, its influence may be secondary to emotional and normative appeals. Engaging users' values through narratives, framing, or feedback mechanisms that reinforce personal and social responsibility may be more effective than focusing on functionality alone. Displaying smart home information that taps into user's environmental values, social norms, or moral responsibility can play a key role in shaping sustained pro-environmental action.

**(2) Designing for continuity and habit formation**

Users who already engage in sustainable practices benefit less from one-time informational interventions and more from features that support ongoing behavior. Tools such as automations, personalized goal-setting, and longitudinal feedback can reinforce sustainable habits over time. Embedding nudges into routine interactions with smart home systems can help maintain engagement without requiring constant user attention. The goal is to normalize energy-efficient behavior as a stable, easy-to-follow part of daily life.

**(3) Providing personalized feedback on financial and environmental benefits**

Users in this study expressed strong interest in receiving tailored insights on how their behaviors translate into monetary savings and environmental impact. Smart home systems should incorporate personalized feedback on energy consumption and actionable recommendations that reflect real-time and historical usage patterns. This feedback should be easy to interpret and embedded within the interface, helping users connect everyday decisions with long-term outcomes.

**(4) Providing information support for "green automation defaults"**

We define "green automation defaults" as pre-configured smart home automations that initiate energy-efficient actions based on specific signals (e.g., grid carbon intensity, electricity price, occupancy). Similar to "green defaults" (Sunstein, 2021; Neumann and Mehlkop, 2020), green automation defaults aim to promote sustainable choices with minimal user effort. Examples include a dishwasher starting when renewable energy availability on the grid is highest, or an electric vehicle beginning to charge when dynamic tariffs enter a low-price period. Given the strong correlation between electricity price and carbon intensity, shifting consumption to low-price periods can also reduce emissions by aligning usage with off-peak, low-carbon hours. Smart home systems are increasingly capable of supporting these automations by using real-time external signals as triggers (Zharova et al., 2024a). Future research should focus on providing informational support for green automation defaults to facilitate sustainable behavior by embedding low-carbon choices into automated household routines.

**(5) Implementing policy mechanisms to financially reward household energy savings**

Survey participants consistently ranked financial incentives, such as tax credits or rebates, as key motivators for engaging in energy-efficient behavior. Policymakers can harness this by developing programs that reward measurable reductions in household electricity consumption (e.g. verified through smart meter or smart home data). These incentives not only reduce the upfront barrier to adoption but also reinforce sustained engagement with smart home technologies for energy efficiency. Aligning behavioral incentives with environmental policy goals presents a powerful pathway for accelerating the energy transition at the household level.

### 5.3 Limitations and future research

This study has several limitations. First, survey participants provided only self-reported behavior. Although quality checks were implemented, it cannot be guaranteed that all respondents were actual smart home users or that all responses were fully truthful. Second, the survey was only available in German and English, which excluded individuals who speak other languages. Additionally, geographical differences were not examined in this study in order to maintain the minimum required sample size for data analysis. Lastly, structural equation modeling can only test relationships that are explicitly specified, meaning that other variables not included in the model may better predict behavioral intention.

Future research should include long-term field studies to examine how smart home user perceptions influence both intentions and actual energy-related behaviors over time, thereby validating the present findings in real-world settings. While behavioral intention is considered a strong predictor of actual behavior, this relationship should be further explored. Research should also investigate the perceptions of user groups beyond early adopters to identify differences that could inform more inclusive smart home designs. Segmentation of user groups may help tailor systems to diverse needs and accommodate changing dynamics within households. Finally, future studies should also assess whether these findings hold across different regions.

## 6 Conclusions

This study provides empirical evidence that personal norms, shaped by environmental awareness and a sense of responsibility, are the strongest predictors of users' behavioral intention to use smart homes for energy efficiency. Subjective norms and attitudes also show statistically significant effects, indicating the importance of social influence and positive perceptions. In contrast, perceived behavioral control has only a minimal effect. These findings suggest that energy-saving behavior in the context of smart homes is driven less by users’ technical abilities and more by their values, motivations, and social environment. To unlock the full potential of smart homes for energy efficiency, system design must go beyond technical functionality. Aligning system features with users’ values, such as providing personalized feedback, supporting automation, and offering targeted incentives, can help close the gap between intention and action.

## Acknowledgements

Alona Zharova greatly appreciates the support of the Joachim Herz Stiftung.

## Appendix A – User Perception and User Behavior: Theories and Models

Summarized results of the comprehensive literature review on theories and models that help explain the relationships between user perception and intention to use Smart Home systems for energy efficiency:

| Theory/Model | Originating Author(s) |
|---|---|
| Theory of planned behavior | Ajzen (1985, 1991) |
| Technology acceptance model | Davis (1986,1989) |
| Unified theory of acceptance and use of technology | Venkatesh et al. (2003) |
| Theory of reasoned action | Fishbein (1967), Ajzen and Fishbein (1973, 1975) |
| Uses and gratification theory | Luo (2002) |
| Health belief model | Rosenstock et al. (1950s) |
| Self-determination theory | Deci and Ryan (1980) |
| Expectation confirmation theory | Oliver (1977, 1980) |
| Mental models | Rouse and Morris (1986), Johnson-Laird (1989) |
| Prospect theory | Kahneman and Tversky (1979) |
| Stimulus-organism-response model | Mehrabian & Russell (1974) |
| Norm activation model | Schwartz (1977) |
| COM-B model | Michie et al. (2011) |
| Transtheoretical model | Prochaska and Velicer (1997) |
| Innovation diffusion theory | Rogers (1962) |
| Social exchange theory | Homans (1958) |
| Attribution theory | Heider (1958), Weiner (1974, 1995) |
| IS-continuance theory | Bhattacherjee (2001) |
| Normalization process theory | May et al. (2009) |
| Simplified abstract model of the human behavior | Aguiar et al. (2015) |
| Uses and gratification expectancy model | Shin (2011) |
| Experiential learning theory | Kolb (1984) |
| Protection motivation theory | Rogers (1975, 1983) |
| Customer satisfaction theory | Oliver (1997) |
| Personal norm theory | Schwartz (1968, 1977) |
| NASSS framework | Greenhalgh et al. (2017) |
| Service-Dominant logic | Vargo and Lusch (2008) |
| Social identity theory | Tajfel & Turner (1986) |
| Socioecological model | McLeroy et al. (1988) |
| Analytic hierarchy process methodology | Saaty (1970) |
| Social representation theory | Moscovici (1984) |
| Concept of animacy | Piaget (1929), Hider & Simmel (1994) |
| Concept of media equation | Reeves & Nass (1998) |
| Social cognitive theory | Bandura (1977) |
| Physical activity for people with a disability model | van der Ploeg et al. (2004) |
| Social cognition model of loneliness | Hawkley and Cacioppo (2010) |
| Privacy calculus model | Culnan and Armstrong (1999) |
| Attachment theory | Bowlby (1969, 1982) |
| Communications privacy management theory | Petronio (2002) |
| Impression management theory | Goffman (1959) |

| Health action process approach | Schwarzer (2008) |
|---|---|
| False consensus theory | Ross et al. (1977) |
| Fuzzy set theory | Zadeh (1965) |
| Persuasive system design model | Oinas-Kukkonen and Harjumaa (2009) |
| Self-regulation framework | Bagozzi (1992) |
| Quality-value-satisfaction-behavioral intention model | Cronin et al. (2000) |
| Self-regulation model of illness | Leventhal et al. (1980), Leventhal et al. (1992) |
| Value belief norm theory | Stern et al. (1999), Stern (2000) |
| Andersen's behavioral model of health services use | Andersen (1995, 2008) |
| Self-perception theory | Bem (1967) |
| IMB framework | Fisher and Fisher (1992), Fisher et al. (2002) |
| Norman's gulf of execution and evaluation | Norman (1986) |
| User competence theory | Huff et al. 1992 |
| Elaboration likelihood model | Petty et al. (1981), Petty and Cacioppo (1986) |
| Social action theory | Ewart (1991) |
| IS success model | DeLone & McLean (1992) |
| Model of news verification behavior | Flanagin & Metzger (2000) |
| Cultivation theory | Gerbner et al. (1994) |
| KISCAP model | Huang (2008) |
| Framework of Dorrestijn | Dorrestijn (2012, 2017) |
| Task technology fit theory | Goodhue and Thompson (1995) |
| Regulatory focus theory | Higgins (1997) |

## Appendix B - Questionnaire

## Understanding User Perception and Intention to Use Smart Home for Energy Efficiency

Thank you for participating in this survey!

With this survey, we would like to investigate the perceptions of people who are currently using smart home systems in their homes and how this influences their energy management and consumption behavior. Understanding how people are using smart home systems can help to improve the existing systems and optimize usability.

In this survey, smart home is defined as a home "… in which data related to a home environment and its residents are obtained from sensors, electric appliances, or home gateway and transferred through a network of communication tools to a monitoring device or execution unit to help decide on or execute proper actions called services. These services are provided either automatically or directly through a remote or central control system to facilitate or improve the residents' daily lives" (Kamel & Memari, 2019). The central smart home hub (also referred to as 'gateway') has a user interface, which can interact with tablets, mobile phones, or computers (Alaa et al., 2017). The smart home system can be connected to various applications, such as air conditioning, heating, ventilation, lighting, security systems, or entertainment devices, amongst others.

Please only participate in this survey if you are a current user of a smart home. There are no right or wrong answers, we are only interested in your personal point of view. This survey will take approximately 10-15 minutes.

The survey is anonymous and cannot be tracked back to you. The data will be used for research purposes.

---

**Smart Home System**

1. Are you currently living in a home that is equipped with a smart home system?*
   [Reminder: In this survey, smart home is defined as a home "… in which data related to a home environment and its residents are obtained from sensors, electric appliances, or home gateway and transferred through a network of communication tools to a monitoring device or execution unit to help decide on or execute proper actions called services. These services are provided either automatically or directly through a remote or central control system to facilitate or improve the residents' daily lives". The central smart home hub (also referred to as 'gateway') has a user interface which can interact with tablets, mobile phones, or computers.]
   ☐Yes ☐ No

2. How long have you been living in a home with a smart home system?
   ☐less than six months ☐6-12 months ☐1-3 years
   ☐4-6 years ☐7-10 years ☐more than 10 years

3. What is the name of the smart home system that you are using?
   [If you're using multiple systems, please list them all in the "other" section]
   ☐Home Assistant ☐Vera ☐OpenHAB ☐Other:____________

4. A smart home energy management system is a system that provides services to monitor and manage electricity generation, storage, and/or consumption in a house, and includes some form of automated planning. (Homan, 2020; Zhou et al., 2016)
   Are you currently using an energy management system as part of your smart home?
   ☐Yes ☐No ☐I don't know.

5. Which products or systems are currently connected to your smart home? Please tick all that apply.

| ☐Smart thermostat | ☐Smart plugs | ☐Smart meter | ☐Heating system |
|---|---|---|---|
| ☐Solar panels (or other energy-generating devices) | ☐Light sources (f.e. smart bulbs) | ☐Air conditioning | ☐Ventilation |
| ☐Water heater | ☐Shading devices | ☐Switches | ☐Sound system |
| ☐Energy storage system | ☐Garage door controls | ☐Door locks | ☐Motion sensors |
| ☐Door and window sensors | ☐Smart TV | ☐Refrigerator | ☐Electric vehicle |
| ☐Stove and/or oven | ☐Dishwasher | ☐Washing machine | ☐Dryer |
| ☐Coffee machine | ☐Microwave | ☐Vacuum cleaner | ☐Mowing robots |
| ☐Fire/smoke/or gas detection | ☐Health-related devices | ☐Security cameras | ☐Smart speaker |
| ☐Streaming devices (e.g. Amazon Fire TV stick, Google Chromecast) | ☐Environmental sensors (such as sensors for temperature, humidity, light) | ☐Flood sensors | ☐Other: |

6. Which person in your household is mainly responsible for your smart home system?
   ☐I am. ☐Responsibility is equally distributed. ☐Another household member. ☐Other:_____

7. Why did you decide to install a smart home system in your home? Please allocate points (0-100) to the listed options depending on their importance to you. All points together must equal 100. Only integer values may be entered in these fields.

| | **Points** |
|---|---|
| Interest in smart home technology | |
| To be more sustainable and save energy | |
| For fun and entertainment purposes | |
| To increase comfort & convenience at home | |
| To save money | |
| To better care for elders or people with disabilities (Ambient assisted living) | |
| To increase safety and security at home | |
| To have more control over my home | |
| It was already installed in the home. | |
| Other: | |
| | **= 100** |

8. From the previous question: What other reasons influenced your decision to install a smart home system?

**Smart Home for Energy Management**

Please indicate how you would rate the following statement.

| [Attitude AT1] | Very unpleasant | Unpleasant | Somewhat unpleasant | Neutral | Somewhat pleasant | Pleasant | Very pleasant |
|---|---|---|---|---|---|---|---|
| I think that **using my smart home for energy management** is/would be … | ☐ | ☐ | ☐ | ☐ | ☐ | ☐ | ☐ |

Please indicate how you would rate the following statement.

| [Attitude AT2] | Very inconvenient | Inconvenient | Somewhat inconvenient | Neutral | Somewhat convenient | Convenient | Very convenient |
|---|---|---|---|---|---|---|---|
| I think that **using my smart home for energy management** is/would be … | ☐ | ☐ | ☐ | ☐ | ☐ | ☐ | ☐ |

Please indicate how you would rate the following statement.

| [Attitude AT3] | Very unsafe | Unsafe | Somewhat unsafe | Neutral | Somewhat safe | Safe | Very safe |
|---|---|---|---|---|---|---|---|
| I think that **using my smart home for energy management** is/would be … | ☐ | ☐ | ☐ | ☐ | ☐ | ☐ | ☐ |

Please indicate how you would rate the following statement.

| [Attitude AT4] | Very unimportant | Unimportant | Somewhat unimportant | Neutral | Somewhat important | Important | Very important |
|---|---|---|---|---|---|---|---|
| I think that **using my smart home to reduce my energy consumption** is … | ☐ | ☐ | ☐ | ☐ | ☐ | ☐ | ☐ |

Please indicate how you would rate the following statement.

| [Attitude AT5] | Very unhelpful | Unhelpful | Somewhat unhelpful | Neutral | Somewhat helpful | Helpful | Very helpful |
|---|---|---|---|---|---|---|---|
| I think that **using my smart** is/would be …**to reduce my energy consumption.** | ☐ | ☐ | ☐ | ☐ | ☐ | ☐ | ☐ |

Please indicate how you would rate the following statements.

| [Attitude AT6, AT7] | Strongly disagree | Disagree | Somewhat disagree | Neutral | Somewhat agree | Agree | Strongly agree |
|---|---|---|---|---|---|---|---|
| I think that **using my smart home helps/would help me to better understand my own energy consumption.** | ☐ | ☐ | ☐ | ☐ | ☐ | ☐ | ☐ |
| My smart home helps/would help me to **save money** by reducing my energy consumption. | ☐ | ☐ | ☐ | ☐ | ☐ | ☐ | ☐ |

## Perceptions about Smart Home for Energy Management

Please indicate how much you agree or disagree with the following statements.

| [Subjective Norm SN1-5] | Strongly disagree | Disagree | Somewhat disagree | Neither agree nor disagree | Somewhat agree | Agree | Strongly agree |
|---|---|---|---|---|---|---|---|
| **Environmental sustainability and climate change** are | ☐ | ☐ | ☐ | ☐ | ☐ | ☐ | ☐ |

| important topics for **people** that are important to me. | | | | | | | |
|---|---|---|---|---|---|---|---|
| **People** who are important to me are trying to **reduce their energy consumption.** | ☐ | ☐ | ☐ | ☐ | ☐ | ☐ | ☐ |
| **People** who are important to me **would approve** of me trying to reduce my energy consumption. | ☐ | ☐ | ☐ | ☐ | ☐ | ☐ | ☐ |
| **People** who are important to me are trying to reduce their energy consumption **by using their smart home.** | ☐ | ☐ | ☐ | ☐ | ☐ | ☐ | ☐ |
| **People** who are important to me think that **I should reduce my energy consumption**. | ☐ | ☐ | ☐ | ☐ | ☐ | ☐ | ☐ |

Please indicate how much you agree or disagree with the following statements.

| [Perceived behavioral control PBC1-6] | Strongly disagree | Disagree | Somewhat disagree | Neither agree nor disagree | Somewhat agree | Agree | Strongly agree |
|---|---|---|---|---|---|---|---|
| I **know how to save energy** with my smart home. | ☐ | ☐ | ☐ | ☐ | ☐ | ☐ | ☐ |
| Controlling my smart home through the **user interface is easy.** | ☐ | ☐ | ☐ | ☐ | ☐ | ☐ | ☐ |
| I have **enough time** to set up and adjust my smart home to reduce my energy consumption. | ☐ | ☐ | ☐ | ☐ | ☐ | ☐ | ☐ |
| I have **enough skills and experience** to reduce my energy consumption with my smart home. | ☐ | ☐ | ☐ | ☐ | ☐ | ☐ | ☐ |
| Reducing my energy consumption is **out of my personal control**. | ☐ | ☐ | ☐ | ☐ | ☐ | ☐ | ☐ |
| **Learning how to use** my smart home for energy management was **simple.** | ☐ | ☐ | ☐ | ☐ | ☐ | ☐ | ☐ |

Please indicate how much you agree or disagree with the following statements.

| [Awareness of consequences AC1-3] | Strongly disagree | Disagree | Somewhat disagree | Neither agree nor disagree | Somewhat agree | Agree | Strongly agree |
|---|---|---|---|---|---|---|---|
| My energy consumption at home **influences the environment.** | ☐ | ☐ | ☐ | ☐ | ☐ | ☐ | ☐ |
| I am **aware of the importance** to reduce energy consumption for environmental reasons. | ☐ | ☐ | ☐ | ☐ | ☐ | ☐ | ☐ |
| I am **concerned about climate change** and its consequences. | ☐ | ☐ | ☐ | ☐ | ☐ | ☐ | ☐ |

Please indicate how much you agree or disagree with the following statements.

| [Ascription of responsibility AR1-3] | Strongly disagree | Disagree | Somewhat disagree | Neither agree nor disagree | Somewhat agree | Agree | Strongly agree |
|---|---|---|---|---|---|---|---|
| **Every individual is responsible** to be more | ☐ | ☐ | ☐ | ☐ | ☐ | ☐ | ☐ |

| mindful about their energy consumption. | | | | | | | |
|---|---|---|---|---|---|---|---|
| **I feel personally obliged** to reduce my energy consumption to the best of my ability, even if it is a small act of my own. | ☐ | ☐ | ☐ | ☐ | ☐ | ☐ | ☐ |
| **I feel guilty** if I unnecessarily increase my energy consumption to improve my own comfort and convenience. | ☐ | ☐ | ☐ | ☐ | ☐ | ☐ | ☐ |

Please indicate how much you agree or disagree with the following statements.

| [Personal norms PN1-3] | Strongly disagree | Disagree | Somewhat disagree | Neither agree nor disagree | Somewhat agree | Agree | Strongly agree |
|---|---|---|---|---|---|---|---|
| **Environmental sustainability** is important to me. | ☐ | ☐ | ☐ | ☐ | ☐ | ☐ | ☐ |
| It is important to me to **focus on reducing my energy consumption** instead of using my smart home for entertainment purposes. | ☐ | ☐ | ☐ | ☐ | ☐ | ☐ | ☐ |
| I feel a **moral obligation** to reduce my energy consumption with the help of my smart home system. | ☐ | ☐ | ☐ | ☐ | ☐ | ☐ | ☐ |

Please indicate how much you agree or disagree with the following statements.

| [Behavioral Intention BI1-5] | Strongly disagree | Disagree | Somewhat disagree | Neither agree nor disagree | Somewhat agree | Agree | Strongly agree |
|---|---|---|---|---|---|---|---|
| I want to continue or start to use my **smart home to manage my energy consumption** in the foreseeable future. | ☐ | ☐ | ☐ | ☐ | ☐ | ☐ | ☐ |
| I **want to (continue to) reduce my energy consumption** with the help of my smart home in the foreseeable future. | ☐ | ☐ | ☐ | ☐ | ☐ | ☐ | ☐ |
| I am **willing to change my energy consumption behavior** to be more environmentally friendly in the foreseeable future. | ☐ | ☐ | ☐ | ☐ | ☐ | ☐ | ☐ |
| I am **willing to sacrifice** some of my comfort to be more environmentally friendly. | ☐ | ☐ | ☐ | ☐ | ☐ | ☐ | ☐ |
| **I want to learn** how to use my smart home for energy efficiency. | ☐ | ☐ | ☐ | ☐ | ☐ | ☐ | ☐ |

## Past Experiences and Motivation

How often have you used your smart home for the following activities on average during the last twelve months?

| [Moderator: Past Behavior PB1-4] | Never | Very rarely | Every other month | Once a month | Every other week | Once a week | Every other day | Daily |
|---|---|---|---|---|---|---|---|---|
| During the last twelve months, I have used my smart home **to monitor my energy consumption**. | ☐ | ☐ | ☐ | ☐ | ☐ | ☐ | ☐ | ☐ |
| During the last twelve months, I have used my smart home to make sure that my **lights or appliances are switched off** when I don't need them. | ☐ | ☐ | ☐ | ☐ | ☐ | ☐ | ☐ | ☐ |
| During the last twelve months, I have **actively tried to reduce my energy consumption** with the help of my smart home. | ☐ | ☐ | ☐ | ☐ | ☐ | ☐ | ☐ | ☐ |
| During the last twelve months, I have used my smart home **to adjust the temperature** when nobody is at home. | ☐ | ☐ | ☐ | ☐ | ☐ | ☐ | ☐ | ☐ |

How likely or unlikely is it that the following smart home services would motivate you to reduce your energy consumption? If you are already using the service, please mark the box on the very right ("Already in use").

| | Very unlikely | Unlikely | Neutral | Likely | Very likely | Alread y in use |
|---|---|---|---|---|---|---|
| Receiving more actionable and easier to understand information on how to save energy at home which is personalized based on my past usage | ☐ | ☐ | ☐ | ☐ | ☐ | ☐ |
| Receiving information on how much energy I consume compared to other people in my neighborhood, friends, or family | ☐ | ☐ | ☐ | ☐ | ☐ | ☐ |
| Setting goals on how much energy I want to use within a specified time frame | ☐ | ☐ | ☐ | ☐ | ☐ | ☐ |
| Receiving information about how much money I could save if I changed my energy consumption behavior | ☐ | ☐ | ☐ | ☐ | ☐ | ☐ |
| Receiving rewards, gifts, or tax deductions depending on how much energy I am saving | ☐ | ☐ | ☐ | ☐ | ☐ | ☐ |
| Receiving information on the effect my energy reduction would have on the environment | ☐ | ☐ | ☐ | ☐ | ☐ | ☐ |
| Having a more fun and game-like smart home user interface | ☐ | ☐ | ☐ | ☐ | ☐ | ☐ |

Have you noticed any cost savings on your energy bills since installing your smart home system?

☐Yes ☐No ☐I don't know.

Since using my smart home, my energy consumption …

☐has increased. ☐has decreased. ☐stayed roughly the same. ☐I don't know.

How would you rate the overall usability of your smart home system and/or smart home energy management system?

☐Very poor ☐Poor ☐Okay ☐Good ☐Very good

**Socio-Demographics**

How old are you? ☐18-24 ☐25-34 ☐35-44 ☐45-54 ☐55-64 ☐65+

What is your gender? ☐Female ☐Male ☐Non-binary/other

What is the highest level of education you have completed?

☐Elementary School ☐Middle School ☐High School ☐Bachelor's Degree

☐Master's Degree ☐Doctoral Degree/advanced degree ☐Other:___________

What is your current occupation status?

☐Employed full-time ☐Employed part-time ☐Self-employed
☐Unemployed ☐Student ☐Retired ☐Other: _________

In which country do you live? _______________

Which of the following best describes where you live?
☐ City ☐Town/semi-dense area ☐Rural area

Are you a … ☐home owner ☐renter ☐other: __________?

How many people live in your household in total (including yourself)?
☐I live alone. ☐Two people ☐3-4 people ☐5-6 people ☐7+

How many of your household members are children (under 18)?
☐0 ☐1 ☐2 ☐3 ☐4 ☐5+ ☐prefer not to say

What is the average monthly net income of your household (after taxes)?
☐less than 1000 USD ☐1001-3000 USD ☐3001-5000 USD ☐5001-8000 USD
☐8001-10000 USD ☐10001-15000 USD ☐15001-20000 USD ☐20001-30000 USD
☐30001-50000 USD ☐more than 50000 USD

Do you have any suggestions for improvements of smart homes and energy management systems, or feedback about the survey, you would like to share?

[Insert Text]

Thank you very much for participating!

## Appendix C – Descriptive Statistics and Survey Results

**Table C1.** Overview of descriptive statistics of respondents.

| Variable | Attribute | Freq. | Prop. |
|---|---|---|---|
| Total (n) | | 363 | 100.0 % |
| Gender | Male | 342 | 94.2% |
| | Female | 19 | 5.2% |
| Age | 18-24 | 15 | 4.1% |
| | 25-34 | 73 | 20.1% |
| | 35-44 | 110 | 30.3% |
| | 45-54 | 93 | 25.6% |
| | 55-64 | 48 | 13.2% |
| | 65+ | 24 | 6.6% |
| Education | High School and below | 135 | 37.2% |
| | Bachelor's degree | 129 | 35.5% |
| | Master's degree | 78 | 21.5% |
| | Doctoral degree/advanced degree | 21 | 5.8% |
| Occupation status | Employed full-time/Self-employed | 304 | 83.7% |
| | Retired | 35 | 9.6% |
| | Student/Unemployed/Part-time | 24 | 6.6% |
| Housing | Homeowner | 296 | 81.5% |
| | Renter | 67 | 18.5% |
| Net household income per month | up to 3,000 USD | 40 | 11.0% |
| | 3,001-5,000 USD | 80 | 22.0% |
| | 5,001-8,000 USD | 93 | 25.6% |
| | 8,001-15,000 USD | 48 | 13.2% |
| | 15,001-50,000 USD | 41 | 11.3% |
| | more than 50,000 USD | 61 | 16.8% |
| Country | Germany | 191 | 52.6% |
| | USA, Canada, UK, Australia | 107 | 29.5% |
| | Other | 65 | 17.9% |

**Table C2**. Product Overview

| | Count | Percentage % |
|---|---|---|
| Smart plugs | 317 | 87.33 |
| Light sources | 306 | 84.30 |
| Switches | 287 | 79.06 |
| Motion sensors | 265 | 73.00 |
| Smart thermostat | 260 | 71.63 |
| Environmental sensors | 257 | 70.80 |
| Smart TV | 240 | 66.12 |
| Door and window sensors | 239 | 65.84 |
| Streaming devices | 228 | 62.81 |
| Smart speaker | 221 | 60.88 |
| Security cameras | 204 | 56.20 |
| Fire, smoke, or gas detection | 158 | 43.53 |
| Shading devices | 134 | 36.91 |
| Sound system | 134 | 36.91 |
| Heating system | 119 | 32.78 |
| Garage door controls | 117 | 32.23 |
| Smart meter | 111 | 30.58 |
| Door locks | 107 | 29.48 |
| Vacuum cleaner | 106 | 29.20 |
| AC | 97 | 26.72 |
| Washing machine | 82 | 22.59 |
| Solar panels | 77 | 21.21 |
| Ventilation | 75 | 20.66 |
| Flood sensors | 72 | 19.83 |
| Water heater | 68 | 18.73 |
| Dryer | 54 | 14.88 |
| Dishwasher | 45 | 12.40 |
| Mowing robot | 38 | 10.47 |
| Refrigerator | 37 | 10.19 |
| Electric vehicle | 37 | 10.19 |
| Health-related devices | 36 | 9.92 |
| Coffee machine | 31 | 8.54 |
| Energy storage system | 30 | 8.26 |
| Stove/oven | 28 | 7.71 |
| Microwave | 10 | 2.75 |

**Table C3.** Ranked importance of factors influencing SH adoption.

| Variable | Average Points (max. 100) | Importance Ranking |
|---|---|---|
| Interest in SH technology | 25.2 | 1 |
| To increase comfort & convenience | 24 | 2 |
| To be more sustainable and save energy | 16.4 | 3 |
| For fun and entertainment purposes | 15.4 | 4 |
| To have more control over my home | 14.1 | 5 |
| To increase safety & security | 13.6 | 6 |
| To save money | 11.5 | 7 |
| Other | 10.8 | 8 |
| Ambient-assisted living (AAL) | 10.3 | 9 |
| Pre-installed | 9.4 | 10 |

**Table C4.** Incentives to use SH for energy efficiency.

| Incentive | Most frequent response | Very likely | % |
|---|---|---|---|
| Rewards | Very likely | 122 | 33.6% |
| Potential financial savings | Very likely | 107 | 29.5% |
| Easier to understand & personalized | Likely | 87 | 24.0% |
| Comparative consumption | Likely | 75 | 20.7% |
| Environmental effect | Likely | 67 | 18.5% |
| Gamification | Neutral | 52 | 14.3% |
| Goal setting | Neutral | 43 | 11.8% |

**Table C5.** Participants' usability rating.

| Usability | Freq. | Prop. |
|---|---|---|
| Very poor | 4 | 1.1% |
| Poor | 18 | 5.0% |
| Okay | 115 | 31.7% |
| Good | 153 | 42.1% |
| Very good | 73 | 20.1% |

## Appendix D – Results

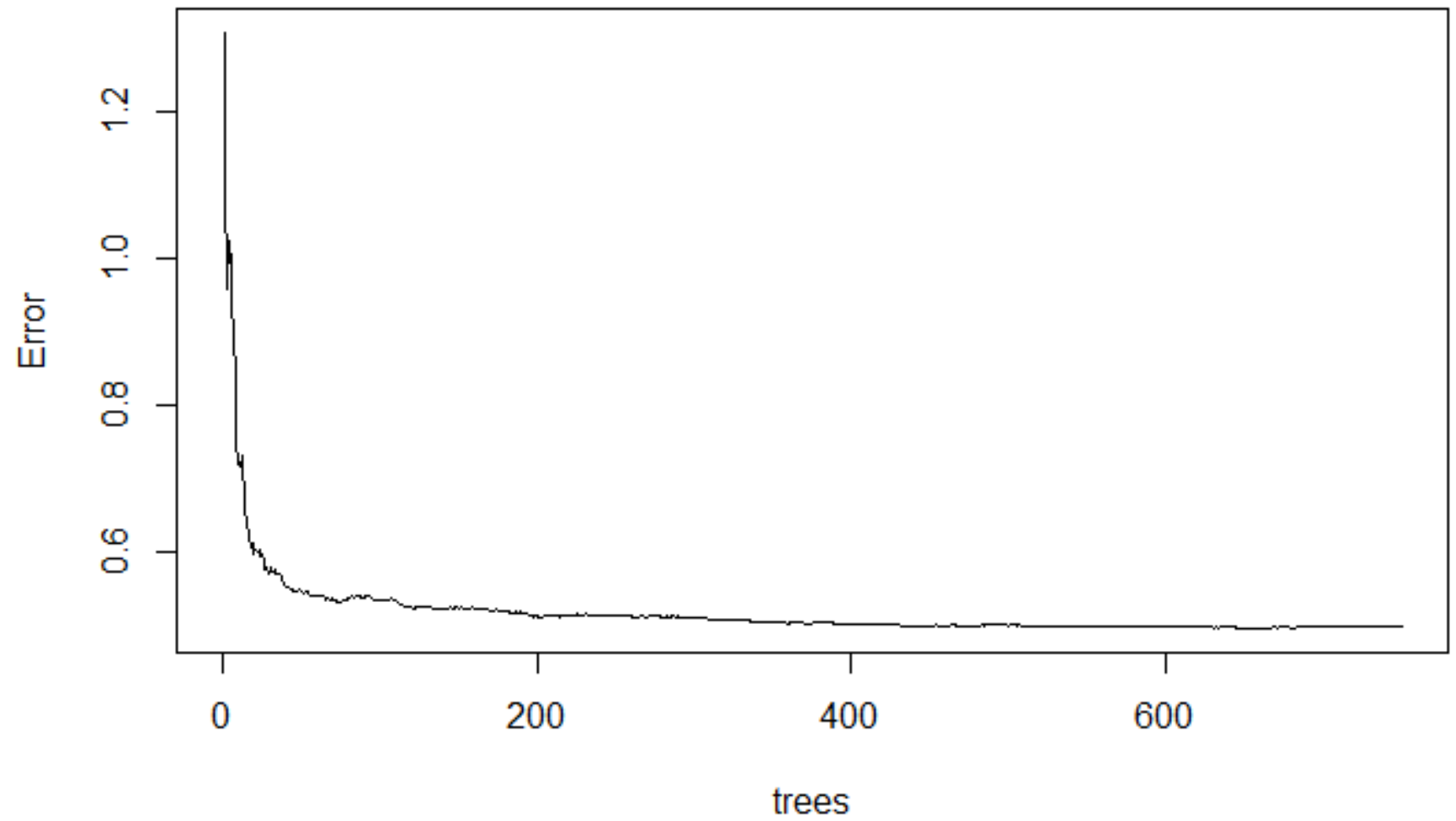

Figure D1. Error over N trees.

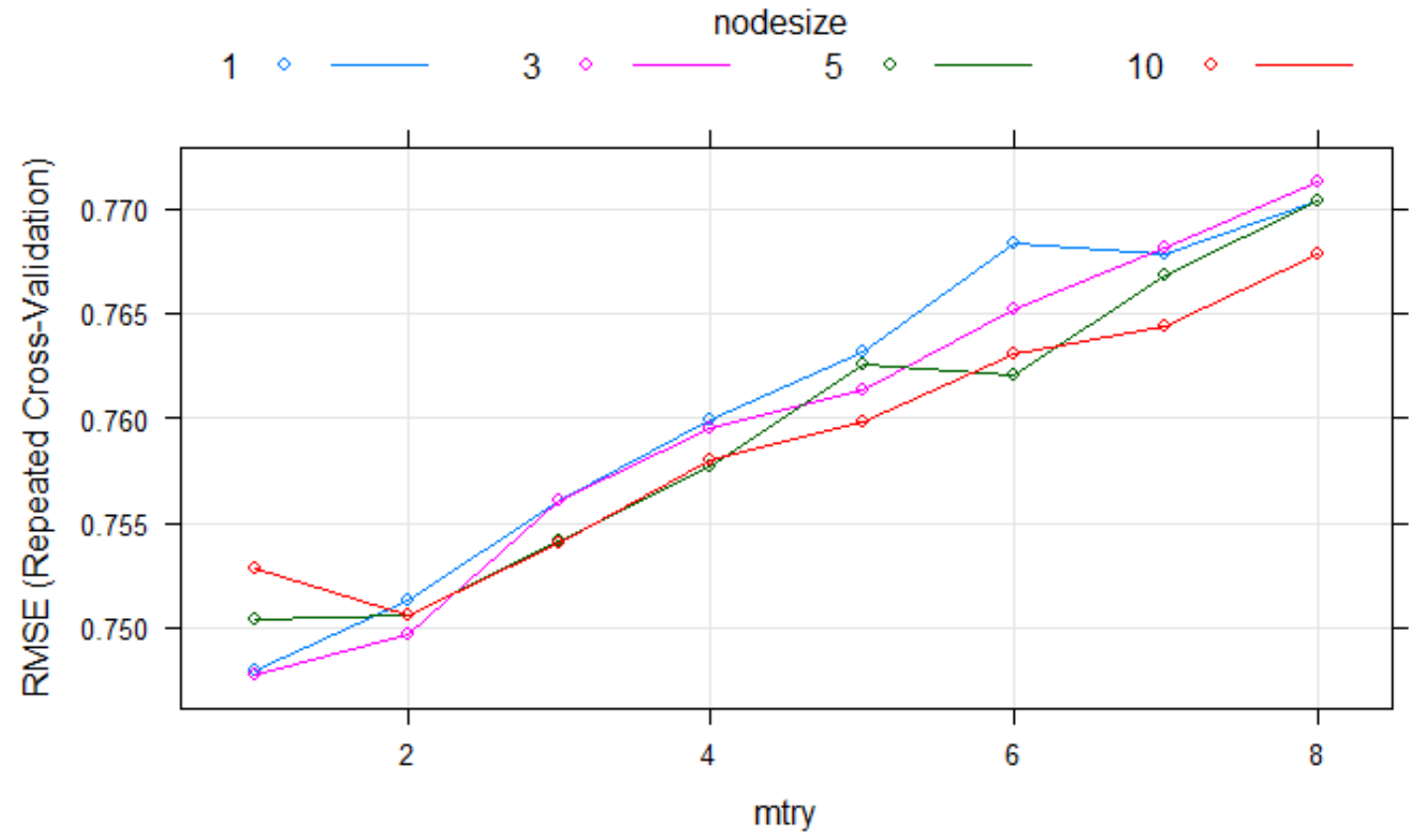

Figure D2. RMSE of Grid Search Optimization Results.